\documentclass[aps,pre,showpacs,twocolumn]{revtex4}

\usepackage{amsmath,amssymb}
\usepackage{graphicx}
\usepackage{color}

\newcommand{\xx}[1]{\textcolor{black}{ #1}}
\newcommand{\xy}[1]{\textcolor{black}{ }}

\newcommand{\dg}{\dot\gamma}

\begin{document}

\title{Rheology near jamming - the influence of  lubrication forces} \author{Moumita
  Maiti} \author{Claus Heussinger} \affiliation{Institute for Theoretical
  Physics, Georg-August University of G\"ottingen, Friedrich-Hund Platz 1, 37077
  G\"ottingen, Germany}

\begin{abstract}
  We study, by computer simulations, the role of different dissipation
  forces on the rheological properties of highly-dense particle-laden
  flows. In particular, we are interested in the close-packing limit
  (jamming) and the question if ``universal'' observables can be
  identified that do not depend on the details of the dissipation
  model. To this end, we define a simplified lubrication force and
  systematically vary the range $h_c$ of this interaction.  For fixed
  $h_c$ a cross-over is seen from a Newtonian flow regime at small
  strain rates to inertia-dominated flow at larger strain rates. The
  same cross-over is observed as a function of the lubrication range
  $h_c$.  At the same time, but only at high densities close to
  jamming, single-particle velocities as well as local density
  distributions are unaffected by changes in the lubrication range --
  they are candidates for ``universal'' behavior.  At
  densities away from jamming, this invariance is lost: short-range
  lubrication forces lead to pronounced particle clustering, while
  longer-ranged lubrication does not.  These findings highlight the
  importance of ``geometric'' packing constraints for particle motion
  -- independent of the specific dissipation model. With the free
  volume vanishing at random-close packing, particle motion is more
  and more constrained by the ever smaller amount of free space. On
  the other side, macroscopic rheological observables, as well as
  higher-order correlation functions retain the variability of the
  underlying dissipation model.
\end{abstract}


\date{\today}



\maketitle

\section{Introduction}

The jamming paradigm aims at providing a unified view for the elastic and
rheological properties of materials as different as foams, emulsions,
suspensions or granular media~\cite{liuNATURE1998}. Structurally, these systems
can all be viewed as dense assemblies of (non-Brownian) particles, and the
particle volume fraction plays the role of the coupling constant that tunes the
distance to the jamming transition. The usefulness of such a unifying concept
hinges on the presence or absence of phenomena that are in some sense
"universal". An important goal then is to delineate these universal aspects of
the jamming transition from system-specific properties that depend on
microscopic details, the driving mechanism or the preparation protocol.

The elasticity and visco-elasticity of jammed packings of particles have
received a lot of attention in the past few years (see e.g. the review
~\cite{hecke2010review}). The elastic moduli and their visco-elastic
generalizations~\cite{PhysRevLett.109.168303} naturally depend on the specific
interaction potential characterising the given particles~\cite{ohern03}.
However, this variation can be understood in terms of a more fundamental -- and
universal -- relation between inter-particle contacts and volume
fraction~\cite{LiuNagelSaarloosWyartREVIEW2010}.

The state of things is much less clear when it comes to flowing systems at
densities close to the jamming transition. A key complication arises because of
the manifold of highly system-specific dissipation mechanisms that characterize
the different systems. Dissipation in emulsions and suspensions is primarily of
hydrodynamic origin, e.g. in the form of lubrication forces or long-range
hydrodynamic interactions. In granular powders, inelastic collisions and dry
friction dominate the dissipation~\cite{hermann}. In wet granular media,
finally, the breaking of liquid capillary bridges between near-by particles is
important~\cite{C2SM25883H}.

Accordingly, the flowcurves (which constitute the appropriate
generalization of the visco-elastic moduli of the packings) display a
wide variety of different forms for different systems. At small
strainrates (or in the hard-particle limit)
Newtonian~\cite{PhysRevLett.109.105901,Lerner27032012} or Bagnold
regimes~\cite{cruz2005PRE,PhysRevLett.109.118305,PhysRevLett.107.188301}
are observed. The associated viscosities diverge at the jamming
transition signalling the impossibility of flow due to packing
constraints. At larger strainrates jamming is usually associated with
shear thinning
\cite{olssonPRL2007,otsukiPRE2009,PhysRevLett.105.175701,PhysRevLett.111.015701}
or possibly with shear-banding when attractive forces are
present~\cite{irani2013condmat}.  When inter-particle friction is
important hysteretic
behavior~\cite{PhysRevE.83.051301,PhysRevE.84.041308,grob2013condmat}
as well as continuous and discontinuous shear
thickening~\cite{heussinger2013condmat,
  denn,PhysRevLett.111.108301,PhysRevLett.100.018301,PhysRevLett.103.086001}
have been observed.

At present it is not clear, however, if there can be any hope to
explain (at least some of) this broad range of rheological properties
on the basis of one or a few generic principles. Progress in this
direction has been made, for example, by defining an underlying
contact network~\cite{Lerner27032012,heussingerPRL2009} in analogy to
what has been achieved for the jammed packings; or by establishing
macro-micro correspondences~\cite{PhysRevLett.108.148301} or scaling
relations~\cite{PhysRevLett.109.105901,PhysRevLett.105.088303,PhysRevLett.109.108001}
between exponents characterizing different aspects of the jamming
singularity.

In this contribution we ask about the influence of different
dissipation mechanisms on the flow properties close to jamming. The
goal is to identify universal observables that do not depend on the
precise nature of the dissipation mechanism, and distinguish them from
system-specific observables that do depend on these details. We expand
on the results of Ref.~\cite{PhysRevLett.109.105901} where a certain
decoupling between the dissipation law and the single particle motion
has been observed. To look into this effect in more detail we will
define a family of dissipation forces and systematically vary the
parameters characterizing this family. With this we can implement a
crossover from a Newtonian to a Bagnold regime and ask about the
sensitivity to these changes on the microscopic level of particle
trajectories.

\section{Simulation}
We study a two-dimensional binary mixture of soft frictionless spheres
of mass $m$. The system of N spheres constitute $\frac{N}{2}$ spheres
of diameter $d$ and $\frac{N}{2}$ spheres of diameter $1.4d$. The
particle volume fraction is defined as $\phi = \frac{\sum_{i=1}^{N}\pi
  R_{i}^2}{L^2}$, where $R_{i}$ is the radius of a particle $i$, and
$L$ is the length of the simulation box. The system is sheared along
the x-direction with a strain rate $\dot{\gamma}$ and Lees-Edwards
periodic boundary conditions are used.

Particles interact via elastic repulsion and via dissipative forces. Two
particles repel each other if they are in overlap with an elastic force,
\begin{eqnarray}\label{eq:}
  \vec{F}_{\rm el} = -\vec{n}_{ij}\epsilon_{ij}(1 - \frac{r_{ij}}{r_c})\,, \qquad r_{ij} < d
\end{eqnarray}
where $\vec{n}_{ij}$ is the unit vector pointing from particle $i$ to particle
$j$, and $r_{ij}$ is the distance between the two particles. The cut-off
$d=(d_i+d_j)/2$ is set by the diameters of the two interacting spheres. In the
small-strainrate limit $\dot{\gamma} \rightarrow 0$ the average overlap vanishes
and the particles effectively behave as hard-spheres ($\epsilon\to\infty$). In
this limit the shear stress is independent of the particle stiffness $\epsilon$.

The dissipative force is modeled as
\begin{equation}\label{eq:diss.force}
  \vec{F}_{\rm diss} = -\zeta \vec{v}_{ij}\,,\quad\mbox{$r_{ij} < d(1+h_c)$}, 
\end{equation}
where $\vec{v}_{ij}=\vec{n_{ij}}[\vec{n_{ij}} \cdot(\vec{v}_i -
  \vec{v}_j)]$ is the relative normal velocity between interacting
particles. The range of the dissipative force $d(1+h_{c})$ is, in
general, taken to be larger than that of the elastic force.  We
characterize this range by the parameter $h_c$, which will be varied
systematically in what follows. \xx{The model is different from
  Durian's bubble model for foams~\cite{durian} mainly due to the
  presence of these lubrication forces: dissipative forces with a
  range longer than the elastic forces. Another difference is that the
  dissipative force in the bubble model is proportional to the total
  relative velocities}


In the limit $h_{c}=0$, particles see each other only when they overlap.
Dissipation is then due to inelastic collisions between particles. This
dissipation mechanism corresponds to the case of a dry granular
powder~\cite{campbell,hermann}. For $h_{c}>0$ particles may interact even without
collision, and Eq.~(\ref{eq:diss.force}) can be viewed as a simplified
lubrication force, where $h_c$ plays the role of the range of the lubrication
interaction. In the following we will therefore call $h_c$ the lubrication
range.

  As discussed our aim here is to understand the role of different
  dissipation interactions on flow properties. To serve this purpose,
  we design a model incorporating the features of suspension flow in a
  simple manner. The model proposed here aims at understanding the
  possible underlying mechanisms of flow in the context of critical
  phenomena at the jamming transition. In real suspensions,
  hydrodynamic forces have a long-range many-body component.  The
  method of Stokesian dynamics is capable of simulating these effects
  efficiently for low and intermediate densities~\cite{brady}. Here,
  we are interested in high densities, close to the maximal density of
  random close packing. In this regime, a common assumption is to
  neglect these long-range hydrodynamic forces and assume them to be
  efficiently screened by the many surrounding particles.  The
  remaining hydrodynamic effects are then lubrication forces, which
  only act when particles are separated by a small gap $h$. For two
  interacting ideal hard-spheres, the leading order term ($h\to 0$)
  can be written in the form of Eq.~(\ref{eq:diss.force}) with a
  distance-dependent viscous coefficient $\zeta(h)$ that diverges upon
  particle collision, $\zeta \sim h^{-1}$~\cite{ballandmelrose}. Next
  to this normal component (squeeze mode) there is also a
  (subdominant) tangential component of the lubrication force (shear
  mode). The tangential viscous coefficient diverges logarithmically
  with the particle gap $\zeta_t(h)\propto -\log(h)$. In reality,
  these divergences are cut-off at small distances set by the surface
  properties of the particles. In general, neither the microscopic
  cut-off nor the range of the lubrication force (it is one term in a
  series) are known.  This warrants a detailed investigation into how
  the functional form of the lubrication interaction affects the flow
  properties at high densities. To this end we provide
  Eq.~(\ref{eq:diss.force}) as a simplified model for the lubrication
  forces.


In most of our simulations we set the \xx{viscous} coefficient to be constant (up to
the range $h_c$) and independent of distance
\begin{eqnarray}\label{eq:zeta.constant}
\zeta(h) = \zeta_0\,,\qquad \xx{h<dh_c}
\end{eqnarray}
Alternatively, we have used
\begin{eqnarray}\label{eq:zeta.exponential}
  \zeta(h) &=& \zeta_0
  \begin{cases}
  1\,,\qquad r_{ij}< d \\
  e^{-h/(dh_{c})}\,,\qquad r_{ij}\geq d
\end{cases}
\end{eqnarray}
\xx{with a cut-off at a large value ($2.5d$)}. It will turn out,
however, that these different choices do not affect the rheological
properties of the system.

With the forces given above Newton's equations of motion
$m\vec{\ddot{r}} = \vec{F^{el}} + \vec{F^{visc}}$ are integrated with
a molecular dynamics (md) simulation using LAMMPS~\cite{link}.  We use
md timestep $\Delta t = 0.01$ and the viscous drag $\zeta_{0} =
2$. The system size $N=1000$ is considered for most of the analysis
unless larger system sizes are required. All quantities are expressed
in units of $d$, $\epsilon$, and $m$ \xx{($d=1$, $m=1$, $\epsilon =
  1$)}.

\section{Results}

\subsection{Close to jamming: $\phi=0.83$}
\begin{figure}[htbp] 
   \centering
   \includegraphics[width=3.5in]{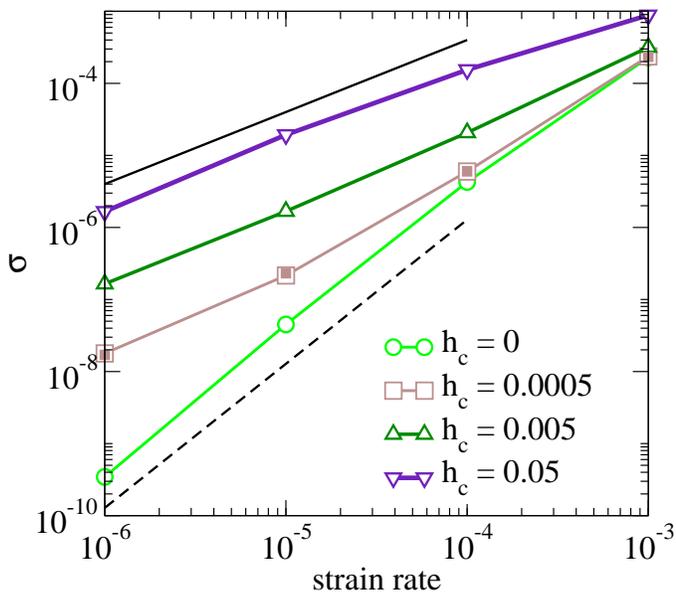}
   \caption{Flow curves $\sigma(\dg)$ at the volume fraction $\phi=0.83$. Thin
     (black) lines represent power-laws $\sigma \propto \dg^{\nu}$ with
     exponents $\nu = 2$ (dashed) and $\nu = 1$ (solid). 
}
   \label{fig:fig1}
\end{figure}

The rheology of a fluid is primarily described by its flowcurve,
i.e. the relation between \xx{shear} stress $\sigma$ and strainrate
$\dg$. In what follows, we fix the volume-fraction to $\phi = 0.83$,
which is close to the jamming transition at $\phi_{\rm rcp} = 0.843$
(random-close packing). \xx{The shear} stress is calculated using
$\sigma = \sum_{i}F_{iy}x_i/A$, where $F_{iy}$ is $y$-component of the
\xx{total force $F = F_{el} + F_{diss}$}, $A$ is area, and it has
dominant conribution from the elastic forces. The particle overlaps
that generate these elastic forces are always at least an order of
magnitude smaller than the lubrication range. Fig.~\ref{fig:fig1}
shows the flowcurves $\sigma(\dg)$ for different lubrication ranges
$h_c\in[0,\,0.05]$.

At the special value $h_{c} =5\cdot10^{-4}$ the data is displayed for
the two variants of the dissipation model, with $\zeta$ taken from
Eq.~(\ref{eq:zeta.constant}) or Eq.~(\ref{eq:zeta.exponential}) (open
and closed squares). As can readily be seen, the flowcurves are
insensitive to this change and therefore independent of the specific
functional form for $\zeta(h_{c})$. As this is also true for the other
quantities we will discuss, in the following we only show data with
$\zeta$ taken from Eq.(\ref{eq:zeta.constant}).

The flowcurves do depend, though, on the \emph{range} of the
dissipation force.  For $h_{c}=0$ (circles) the shear stress scales
quadratically with strainrate, $\sigma = \hat\eta\dg^2$. In this
limit, the dissipative force Eq.~(\ref{eq:diss.force}) only acts when
particles overlap, i.e. during a particle collision. This regime is
well-known from the flow of granular
powders~\cite{campbell,hermann}. The quadratic (``Bagnold'') scaling
with strainrate can readily be explained by dimensional
analysis~\cite{lemaitrepre}.  Shear stress (in two dimensions) is work
done per unit area and the dimensional analysis gives $\sigma \sim
m\dg^2$, i.e. the effective viscosity is proportional to particle
mass, $\hat\eta \sim m$.  Hence, rheology in the limit $h_{c}=0$ is
ascribed to inertia (``inertial regime'').

Increasing $h_c$ from zero, there is a crossover to a second,
``viscous regime'', where, $\sigma = \eta \dg$, a Newtonian
liquid. Here viscous lubrication forces dominate over inertial forces,
and dimensional analysis gives $\eta\propto \zeta_0$. The crossover is
also seen at a fixed $h_c$, from viscous regime at low strainrate to
inertial regime at higher strainrate~\footnote{The slight leveling off
  in the flowcurves at large strainrates $\dg\approx 10^{-3}$ is due
  to the finite stiffness of the particles $\epsilon<\infty$.  This
  will not be discussed here.}. A similar crossover is also present in
experiments~\cite{coussot,fall}, \xx{and recent
  simulations~\cite{vagberg}}. It may also be interpreted in terms of
shear-thickening as a consequence of inertial effects, similar to what
has been seen in Ref.~\cite{PhysRevLett.111.108301}.

\begin{figure}[htbp] 
\centering
\includegraphics[width=3.5in]{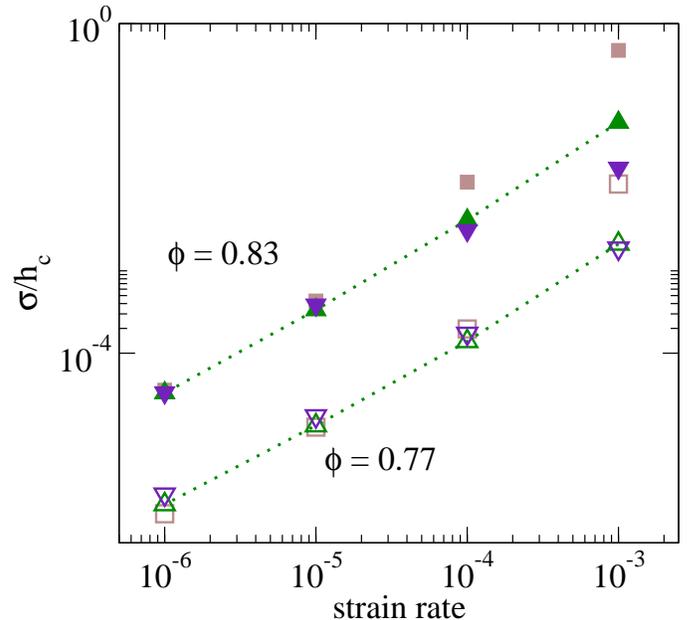}
\caption{Flow curves normalized by lubrication range $h_c$ at densities 0.83 and
  0.77. The coloring is the same as in Fig.~\ref{fig:fig1}.}
\label{fig:fig2}
\end{figure} 

Fig.~\ref{fig:fig2} shows that the stress $\sigma\propto h_c$ in the
viscous regime. This scaling can be understood from an energy
  balance between injected work and dissipated energy. The key
  ingredient is the $h_c$-dependent dissipation volume (area) $v_{d} =
  (\pi d^2)((1+h_c)^2-1)\sim \pi d^2 h_c$ ($h_c$ is
  dimensionless!). The scaling expression for the energy balance
  between work and dissipation reads: $d^2\sigma\dg \sim
  \zeta\dg^2v_d$, where $\sigma\dg$ represents the density of work and
  $\zeta\dg^2$ the dissipation density.
  The stress then follows as $\sigma\sim\zeta\dg h_c$. Hence, the
  scaling $\sigma\propto h_c$.


Looking in more detail, the energy balance equation can be written as
$\sigma\dg \propto 2\pi\zeta\int dh(d+h) \,g(h)v^2(h)$, where $v^2(h)=\langle
v_{ij}^2\rangle_h$ is the normal component of the relative velocity of
two particles $i$ and $j$ (as in Eq.~(\ref{eq:diss.force})) and the
average is taken over all interacting particle pairs at a prescribed
gap $h$. Furthermore, $g(h)$ is the pair-correlation function.

\begin{figure}[tbp]
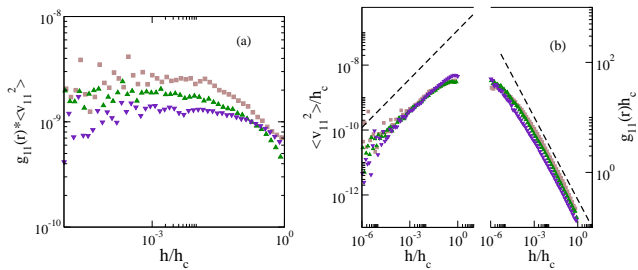
 
  \centering \includegraphics[width=1.5in, clip=true]{fig3a.eps}
  \includegraphics[width=1.75in, clip=true]{fig3b.eps}
  \caption{(a) Average dissipation density vs rescaled gap
    $h/h_c\equiv (r-\sigma)/(dh_c)$ for different $h_c$ (only
    small-small ($11$) particle pairs are accounted for). (b) Rescaled
    relative velocities and pair-correlation function vs. gap
    $h/h_c$. Dashed lines indicate power-law $x^{1/2}$ and $x^{-1/2}$,
    respectively. The coloring is the same as in
    Fig.~\ref{fig:fig1}. \xx{Strain rate is $\dg = 10^{-5}$}.}
  \label{fig:hdependence}
\end{figure}

  As can be seen in Fig.~\ref{fig:hdependence}, the dissipation
  density $\zeta g(h)v^2(h)$ is approximately constant and only weakly
  dependent on $h$ and $h_c$, thus validating the scaling
  ansatz. Notably, the individual contributions $g(h)$ and $v^2(h)$
  are not constant and furthermore strongly depend on the lubrication
  range $h_c$. Interestingly, they both obey approximate scaling forms
  $v^2(h) = h_c{\cal F}(h/h_c) \sim 1/g(h)$ with a scaling function
  ${\cal F}(x)\approx x^{1/2}$. Thus, both relative particle
  velocities and local structure do vary strongly with dissipation
  range while dissipation densities does not.

\begin{figure}[tbp] 
    \centering
  \includegraphics[width=3.0in]{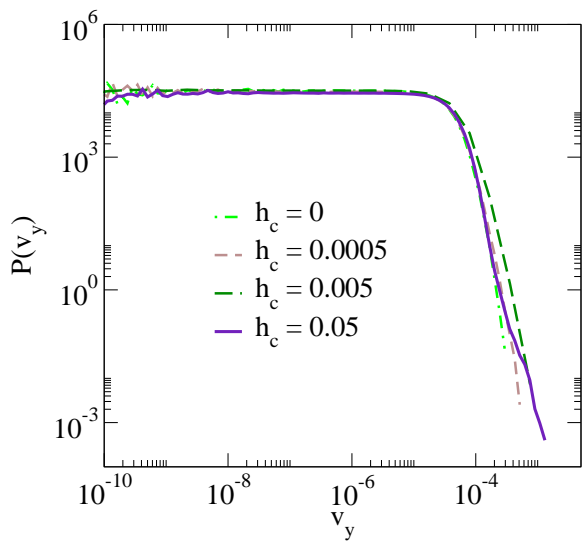}
  \caption{The distribution function of particle velocities $|v_{y}|$ at volume
    fraction $\phi = 0.83$. The strain rate is $\dot{\gamma} = 10^{-5}$.}
  \label{fig:fig3}
\end{figure}

Fig.~\ref{fig:fig3} shows the probability distribution of
\emph{single} particle velocities $v_{y}$ (gradient
component). The velocity component $v_{x}$ (flow direction) has both
affine and non-affine contributions, while the component $v_{y}$ is
purely non-affine. So, we look at the distribution of $v_{y}$ to avoid
mixing of affine and non-affine motion.

Quite surprisingly, and distinct from the relative velocities, the
distribution of single-particle velocities hardly changes by changing
the range $h_c$ of the lubrication force~\footnote{Small variations
  are visible in log-lin representation, especially what regards the
  height of the plateau~\cite{PhysRevLett.109.105901}.}. Recall, that
at the same time, the flowcurve (Fig.~\ref{fig:fig1}) changes by
orders of magnitude and even the functional form $\sigma(\dg)\sim
\dg^x$ changes, from $x=2$ to $x=1$. Apparently, all these changes can
be accomplished with only minimal changes in the statistical
properties of the single particle velocities. Below, we will
  furthermore see that the local particle density behaves similarly
  (Fig.\ref{fig:fig5}b). Thus, it seems that single-particle
  (one-point) observables, and in particular the single-particle
  velocities are independent of dissipation range $h_c$, while
  higher-order correlations (multi-point observables) are very
  sensitive to details of the dissipation model.


\begin{figure}[htbp] 
  \centering
  \includegraphics[width=3in,clip=true]{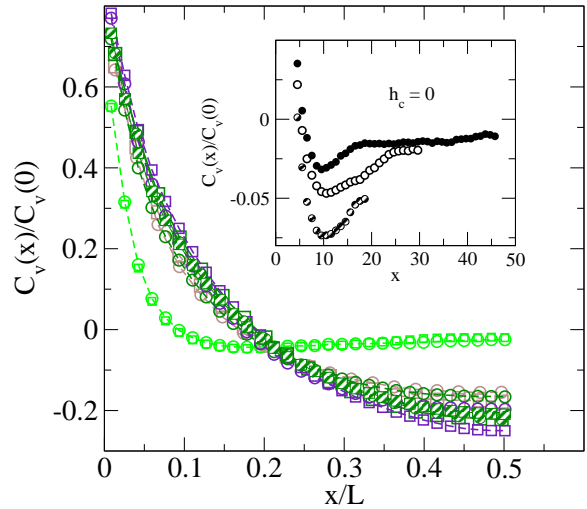}
  \caption{The velocity spatial correlation $C_v(x)/C_v(0)$ (see
    text) at volume fraction $\phi = 0.83$ for two strainrates $\dg =
    10^{-5}$ (circle) and $10^{-6}$ (square). Four cases of different
    $h_c$ values are shown. \xx{Two system sizes $N = 1000$ (striped) and
    $N=2500$ (open) are shown. The
    coloring is the same as in Fig.~\ref{fig:fig1}. The x-axis is
    normalized by box-length $L$. Inset shows without this
    normalization for $h_c = 0$ for three system sizes $N = 1000$ (striped)
    $2500$ (open), $6000$ (filled)}.}
  \label{fig:fig9}
\end{figure}

  To reinforce this finding we look at a velocity-correlation
  function.
  We calculate velocity spatial correlation
  \xx{$\frac{C_v(x)}{C_v(0)}$, where $C_v(dx) = \langle
    v_y(x;y)v_y(x+dx;y)\rangle$ and $C_v(0) = \langle v_y^2\rangle$}
  the normalization factor.
  The relative velocities $v_{ij}$ of interacting particle pairs are
  related to the short-distance part of this function. We are now
  interested in the long-distance behavior far beyond the interaction
  cut-off.
  Fig.~\ref{fig:fig9} clearly shows that the correlation function
  changes its behavior qualitatively, when $h_c$ is increased from
  zero to finite values. For $h_c=0$ it has a minimum at a certain
  distance that is independent of system-size (see inset).
  With lubrication interactions the minimum quickly disappears and the
  correlation function rather monotonically decays with a length-scale
  that depends on the system-size.
%
  Quite similar results on the correlation function have been reported
  in a related system~\cite{PhysRevLett.109.168303}. There, two
  different dissipation forces have been compared in the context of
  Durian's bubble model~\cite{durian}.



 These findings throw an interesting light on the role of
  dissipative and elastic forces in this system.
  Recall, that the stress is independent of particle stiffness
  $\epsilon$, i.e. particles effectively behave as hard spheres
  ($\epsilon\to\infty$): the elastic forces then very effectively
  serve to reenforce the volume exclusion between nearby
  particles. The free volume of a hard-sphere system vanishes at
  random-close packing, thus, particle motion is more and more
  constrained by the ever smaller amount of free space, as random
  close packing is approached from below. It is this geometric
  ``singularity'' that provides strong constraints for particle
  motion. In previous work~\cite{heussingerEPL2010}, we have shown how
  single-particle velocities $\langle v_y^2\rangle$, as a result of
  these constraints, actually \emph{diverge}, when RCP is approached.
Such a geometric mechanism is independent of the specific dissipative
force, and therefore of the dissipation range $h_c$, as observed in
Fig.\ref{fig:fig3}. The remaining role of dissipation then is to
determine the higher-order contributions to the particle motion, and
indeed, the \emph{amount of dissipated energy} along geometrically
predetermined trajectories.


Naturally, this picture only works, if the steric effects of
  volume exclusion provide strong constraints on particle
  motion. Thus, the closer to random-close packing the better. This
  was shown explicitly in Ref.~\cite{PhysRevLett.109.105901}, where
  the critical density was approached up to $\phi_{\rm rcp}-\phi =
  0.003$. In the following section we will see how by \emph{reducing}
the density and going away from close packing, the newly available
space can be used.

\subsection{Away from jamming: $\phi=0.77$}

\begin{figure}[htbp]
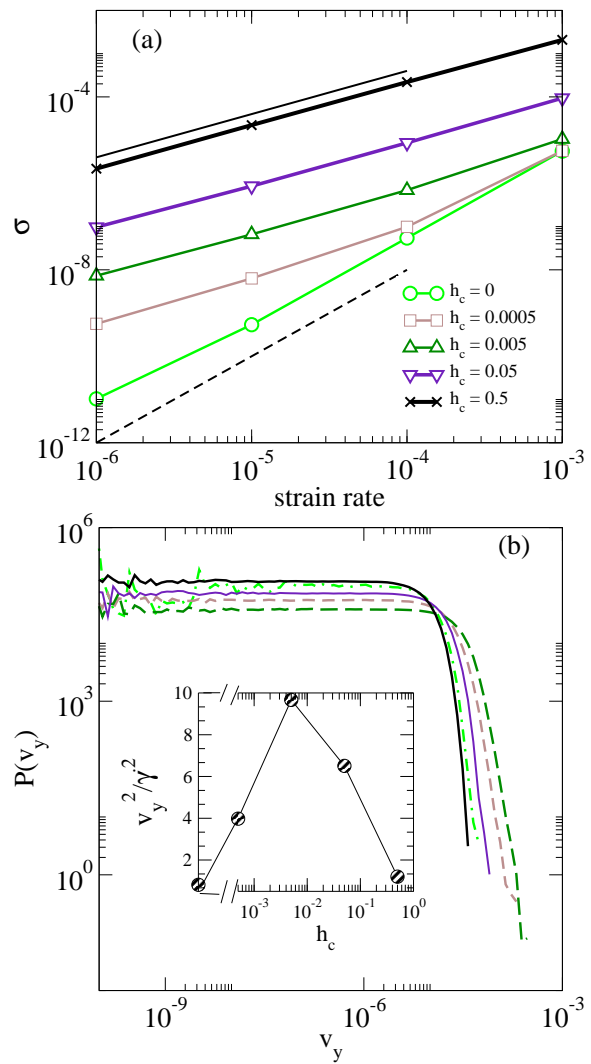
 
\centering
\includegraphics[width=3.0in]{fig6a.eps}
\includegraphics[width=3.0in]{fig6b.eps}
\caption{(a) Flowcurves for volume fraction $\phi = 0.77$. (b) The distribution
  function of $|v_{y}|$ at volume fraction $\phi = 0.77$.  Coloring is the same as
  Fig.\ref{fig:fig1} (in addition, data for $h_{c} = 0.5$ (black line and cross
  symbol) is shown). Strain rate is $\dg = 10^{-5}$. Inset in panel (b) shows
  the second moment $\langle v_{y}^2\rangle $ vs. lubrication range $h_{c}$,
  where $y-$axis is divided by $\dg^2$.}
\label{fig:fig4}
\end{figure}

We fix the packing-fraction $\phi = 0.77$ away from random close
packing.  Fig.~\ref{fig:fig4}(a),(b) show the flowcurves and the
associated single-particle velocity distributions. Like at $\phi =
0.83$, the same crossover between inertial and viscous regime is
observed.  Also the scaling with $h_c$ in Fig.~\ref{fig:fig2} is
  present with small deviations.  Unlike at $\phi = 0.83$, however,
the velocity probability distribution function \emph{is} changing with
$h_c$ (Fig.~\ref{fig:fig4}(b)) .

This  highlights that the
single-particle motion is no longer governed only by the geometry of
packings. Particles now have enough space to move. The geometric
  constraints are relaxed and particle motion is a result of a complex
  interplay of elastic and dissipative forces.

Looking more closer into the variation of the velocity distribution function in
Fig.~\ref{fig:fig4}(b), one observes that the distribution function is
non-monotonic with the lubrication range $h_c$. The tail is enhanced at small
$h_c$, indicating faster movement of particles. Interestingly by further
increasing $h_c$ this enhancement goes away and particle velocities are almost
the same as at $h_{c}=0$. This trend is clearly visible in the second moment of the
distribution ($\langle v_{y}^2\rangle$), which we show in the inset of
Fig.~\ref{fig:fig4}(b). Indeed, typical non-affine velocities are maximal at
intermediate values of the lubrication range.

\begin{figure}[htbp] 
  \centering
  \includegraphics[width=1.5in,angle=-90]{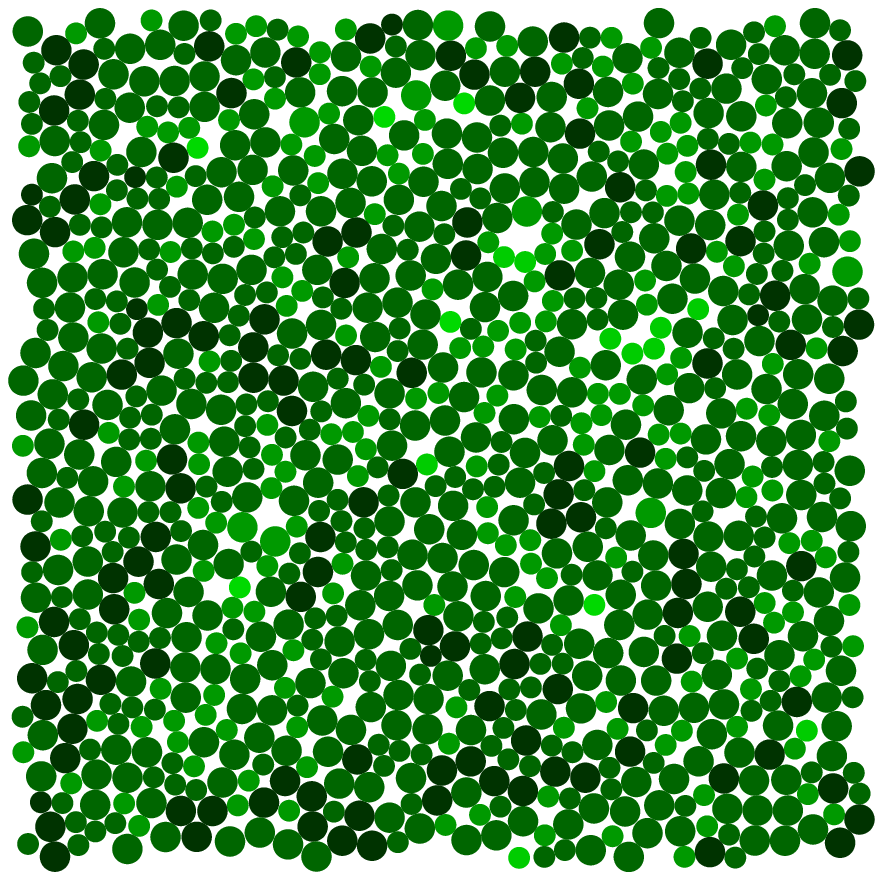}
  \includegraphics[width=1.5in,angle=-90]{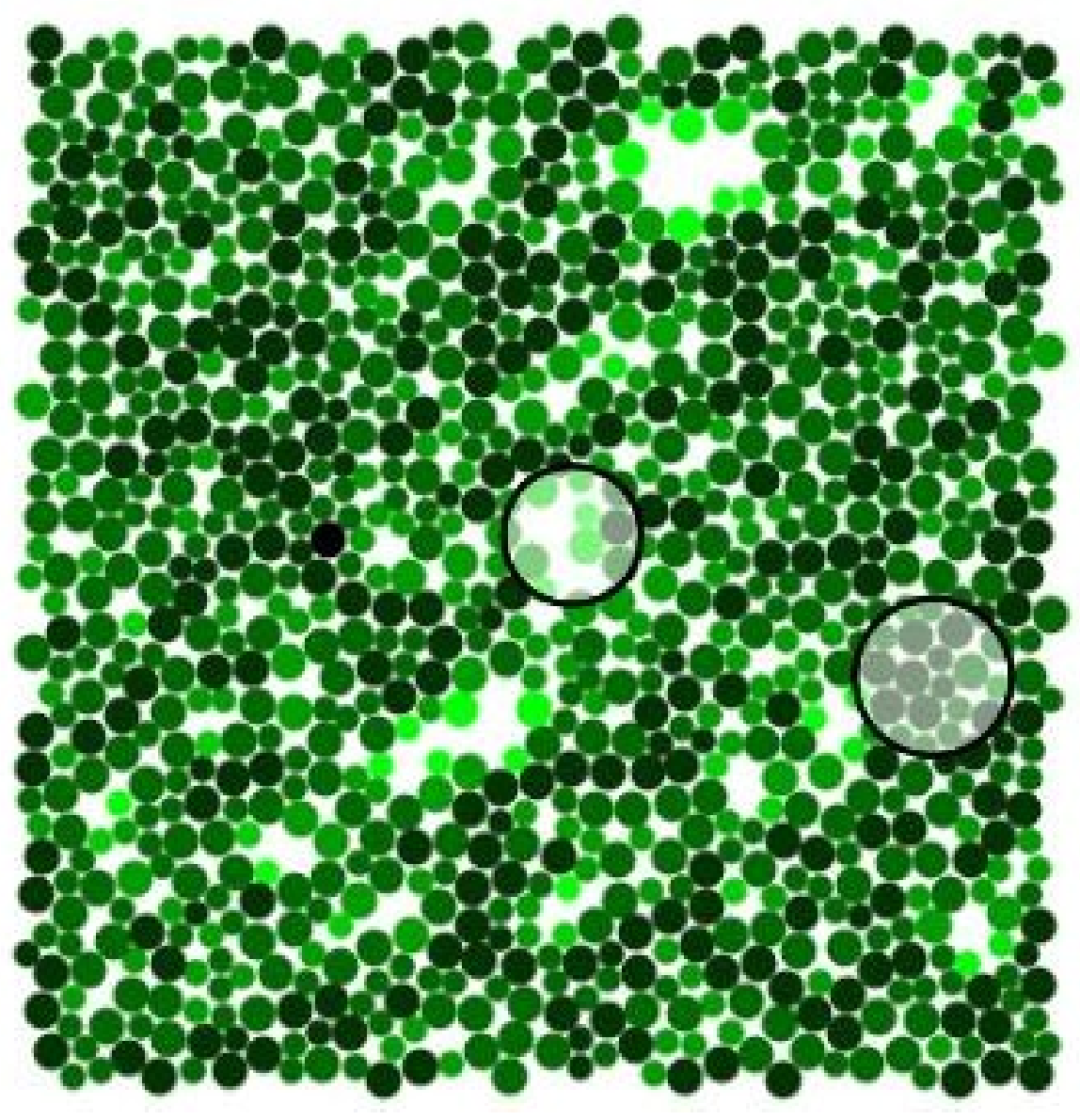}
  \caption{Snapshots for (a) $h_{c} = 0$ and (b) $h_{c} =
    0.005$. Volume fraction $\phi = 0.77$.  Coloring is done from
    high(dark) to low(light) local density. Possible clusters or
    `holes' are highlighted by circles.}
  \label{fig:fig6}
\end{figure}

There may be a change in structure coupled to this velocity maximum so we look
at snapshots. Fig.~\ref{fig:fig6} shows two snapshots one at $h_{c}=0$, the other at
$h = 5\cdot 10^{-4}$. The structure at $h_{c}=0$ is homogeneous whereas at $h_{c} =
5\cdot 10^{-4}$ it displays density fluctuations and particle clustering.

\begin{figure}[htbp]
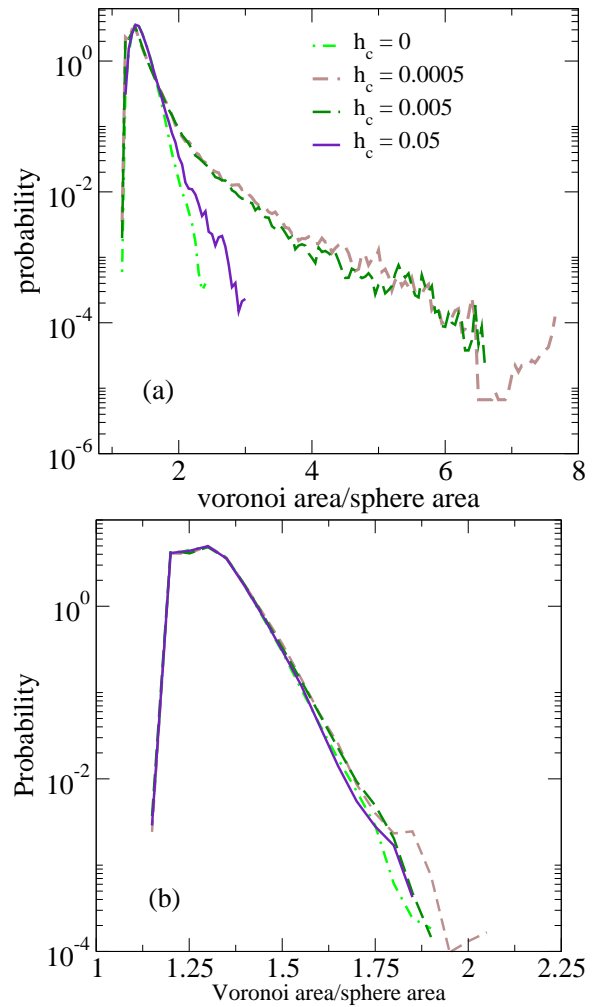
 
  \centering
  \includegraphics[width=3.0in]{fig8a.eps}
  \includegraphics[width=3.0in]{fig8b.eps}
  \caption{The probability distribution of local Voronoi area (inverse local
    density) at $\dot{\gamma} = 10^{-5}$.  (a) $\phi = 0.77$, (b) $\phi = 0.83$.
    The Voronoi area is normalised by respective sphere area.}
  \label{fig:fig5}
\end{figure}

We have quantified these density fluctuations by calculating local
Voronoi area using a ``generalized Voronoi''
algorithm~\cite{maiti}. The density is large if the area is small and
{\it vice versa}. Fig.~\ref{fig:fig5}(a) shows the probability
distributions of local Voronoi area of the smaller particle for
different lubrication ranges $h_c$. We show data of one component as
the distribution is qualitatively similar for both particle sizes,
except the peak position is different. A broad tail is observed for
small $h_c$ which is a clear indication of large voids and associated
particle clustering. The tail comes back to normal ($h_{c}=0$)
behavior by further increasing $h_c$.  This implies that clustering
disappears by increasing $h_c$. The distribution of local density is
therefore non-monotonic, and behaves much in the same way as the
velocity distribution function Fig.~\ref{fig:fig4}(b).

  Lubrication-induced clustering should be taken as a consequence of
  the minimization of energy dissipation. Indeed, forming clusters is
  expected to reduce the relative motion of nearby particles, thus
  reducing their dissipation.  However, we have also seen that
  particle clustering comes at the cost of enhanced single-particle
  motion (Fig.\ref{fig:fig4}(b)).  This may seem paradoxical, at
  first. However, previous work on related systems have shown how
  clusters can be viewed as particles with renormalized diameter
  $\tilde d$~\cite{PhysRevLett.109.105901,heussinger2013condmat}. The
  velocity scale of such a particle is $v\sim \dg\tilde d$ (on
  dimensional grounds), and thus increases quite naturally with the
  size of the particle/cluster. With the data taken from the inset of
  Fig.~\ref{fig:fig4}(b) we can estimate the cluster size('hole'
    size) to be on the order of $\tilde d \sim 3$. While this seems
  to be in rough agreement with the size of the inhomogeneities seen
  in Fig.~\ref{fig:fig6}, care must be taken and cluster size should
  be measured directly to quantitatively confirm this speculation.

Finally, note that at $\phi = 0.83$ the distribution of Voronoi areas is nearly
independent of the lubrication range $h_c$ (Fig.~\ref{fig:fig5}(b)). A small
enhancement of the tail can be seen indicating a very weak clustering also at
the higher density. This may explain the weak dependence of the velocity
distribution function on $h_c$ as seen in Fig.~\ref{fig:fig3}. This effect is
expected to vanish upon increasing the density further towards random-close
packing.

To conclude, at densities away from the close-packing threshold the
motion of particles is governed not only by steric constraints but by
a complex interplay of conservative and dissipative forces.  In
particular, we observe that short-range lubrication forces induce
particle clustering.  Longer-range lubrication does not lead to
clustering. Rather several observables, like the local density or the
velocity distribution, have much the same form as without any
lubrication forces (i.e.  with only inelastic collisions as
dissipation mechanism).

\section{Discussion and Conclusion}

The goal of this work was to investigate the role of dissipative forces for the
rheological properties of highly dense particulate flows. The question is if and
under what circumstances ``universal'' observables can be identified that do not
depend on the details of the dissipative model. To answer this question, we
defined a simplified lubrication force between near-by particles, and
systematically varied the range $h_c$ of this interaction.

On the level of the flowcurves we observe a transition from a Bagnold
to a Newtonian regime by changing either $\dg$ or $h_c$. The stress
can be written as
\begin{equation}
\sigma =   \eta_Nh_c\dg+\eta_B\dg^2 
\end{equation}
with volume-fraction dependent ``viscosities'' $\eta_N$ and
$\eta_B$. While from our simulations we cannot make any definite
statement about the volume-fraction dependence, from previous work one
expects a power-law divergence at the close-packing threshold, thus
$\eta_N\sim(\phi_{\rm rcp}-\phi)^{-\beta}$ and $\eta_B\sim(\phi_{\rm
  rcp}-\phi)^{-\alpha}$ with $\alpha\approx 4$ and $\beta\approx
2$~\cite{otsukiPRE2009,PhysRevLett.109.105901}. This entails a
crossover strainrate $\dg_c \sim \delta\phi^{\alpha-\beta}h_c$ that is
vanishing at $\phi_c$.
The crossover stress $\sigma_c\sim\delta\phi^{\alpha-2\beta}$ is
expected to be only weakly density dependent.  Note, that
qualitatively similar results have been found in the experiments of
Ref.~\cite{fall}, albeit over an extended volume-fraction range, where
the simple power-law dependence is not expected to hold anymore.


Going beyond the characterization of the flowcurves we found that
systems at random-close packing are only weakly sensitive to the
lubrication range. In particular, the single-particle velocity and the
local density distributions are nearly unchanged by varying $h_c$ and
therefore serve as candidates for ``universal'' behavior. We
attribute this invariance to the strong geometrical constraints of
excluded volume which do not allow particles to react to changes in
the dissipative force.

\begin{figure}[htbp] 
\centering
\includegraphics[width=1.65in]{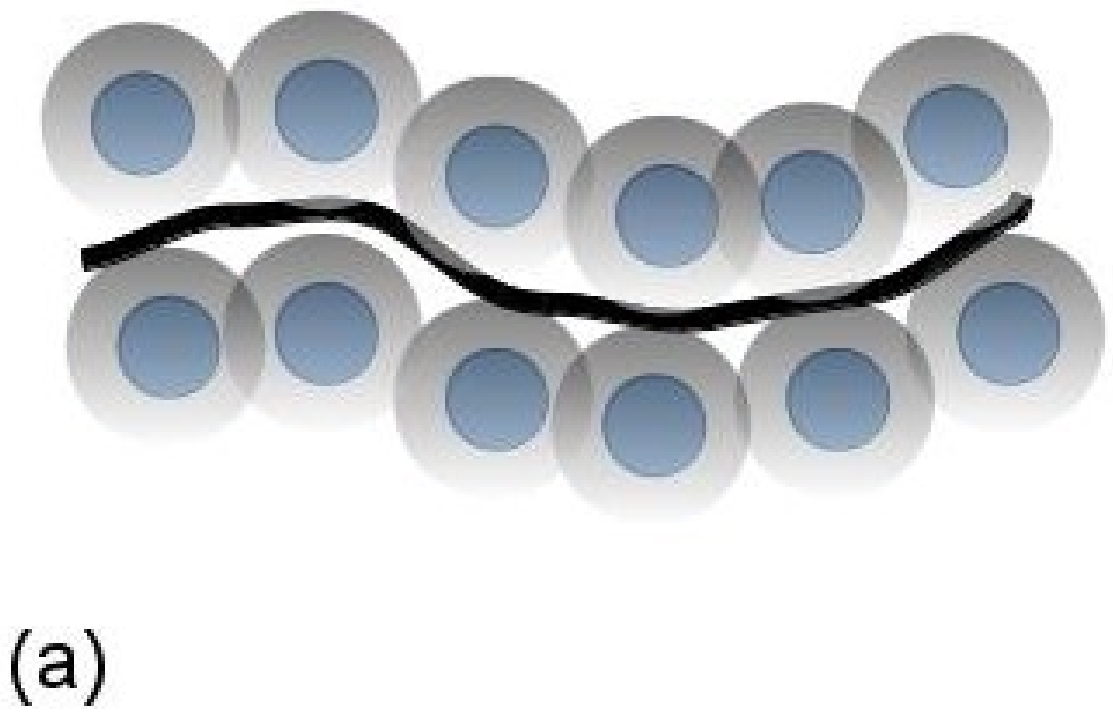}
\includegraphics[width=1.45in]{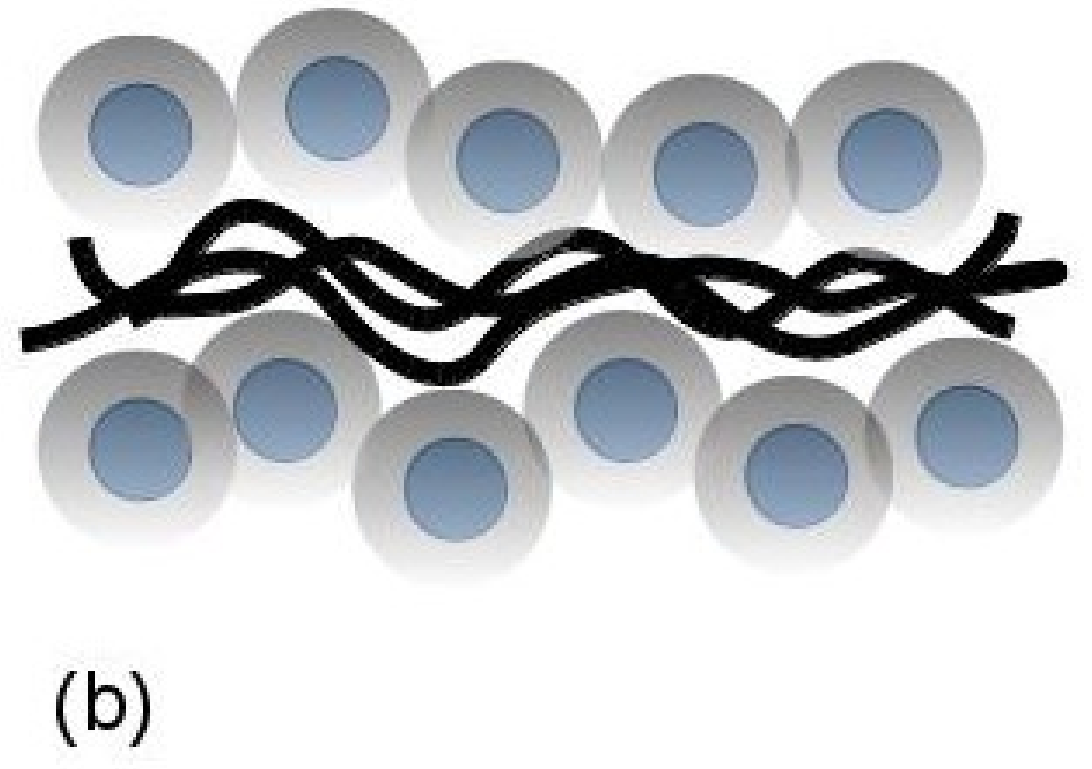}
\caption{Schematic of possible paths arising from geometric aspects. A mobile
  test particle passing through an immobile environment. A high densities the
  narrow channel in (a) very effectively restricts the motion of the mobile
  particle. At lower densities (b) many different passages through the channel
  are possible.  The bigger sized (opaque) circles represent the excluded volume
  for the motion of the mobile test particle.}
\label{fig:figs}
\end{figure}

The simple schematic in Fig. \ref{fig:figs} illustrates the underlying idea.
The geometric constraints of a dense immobile environment force a single mobile
particle to move along a well defined trajectory through a narrow channel.
Dissipative forces only determine the fine-details of the transit encoded, for
example, in the relative particle motion during collisions. More generally, it
seems that changes in the dissipation show up only in higher-order correlation
functions as observed in spatial velocity correlation, but not in the 
single particle motion. Dissipation, then, decides
upon the amount of energy that is necessary to push the particle through the
channel. 



At densities away from close-packing, geometrical constraints are weak and
particles \emph{do} respond to the dissipation mechanism (like in the wide
channel of Fig.\ref{fig:figs}(b) ).  The qualitative features of this response
strongly depend on lubrication range $h_c$. For small $h_c$ we find 
pronounced particle clustering evident, for example, in the distribution of 
local Voronoi volumes.

Particle clustering as a result of lubrication forces is well known in the
literature of dense suspension flows~\cite{wagner}
and are believed to be responsible for the phenomenon of
shear-thickening.  Clusters form because particles that are pushed
together by the flow experience the high \xx{viscous} coefficient of the
lubrication force. Shear thickening then occurs, because smaller and
smaller inter-particle gaps are probed when the strainrate is
increased.
The lubrication force keeps particles together, the repulsive force
drives particles apart. This delicate balance is shifted towards
closer gaps, when strainrate is increased. In our simplified model, no
such balance at finite gaps is possible. As a consequence, particles
either ``fully'' cluster or they don't cluster at all. There is no
effect of strainrate. Rather, we view clustering as a possibility for
the particles to better utilize the free space in order to
\emph{reduce} energy dissipation and thus viscosity (as compared to
the unclustered state).

Finally, we have shown that long-range lubrication forces do not lead to
clustering.  Apparently, long-range lubrication forces are very effectively
screened and only act as a kind of mean field.  It would be tempting to
speculate that the full long-range hydrodynamic interactions share the same fate
at high enough particle densities as our simplified lubrication interactions do.

\begin{acknowledgments}
  We acknowledge financial support by the DFG via the Emmy Noether program (He
  6322/1-1).
\end{acknowledgments}


\end{document}